\documentclass[showpacs,aps]{revtex4}

\usepackage{graphics}
\usepackage{graphicx}
\usepackage{dcolumn}     
\usepackage{bm}          
\usepackage{subfigure}
\usepackage{amsmath}
\usepackage{amssymb}
\usepackage{setspace}
\usepackage{array}
\usepackage{latexsym}
\usepackage[warning,math]{easyeqn}
\usepackage[thinlines]{easybmat}
\newcommand{\ds}{\displaystyle}
\newcommand{\scs}{\scriptscriptstyle}

\begin{document}
\title{The Anti-FPU Problem} \author{Thierry
  Dauxois$^{1}$\thanks{E-mail: Thierry.Dauxois@ens-lyon.fr}, Ramaz
  Khomeriki$^{2}$, Francesco Piazza$^{3}$, Stefano
  Ruffo$^{1,4}$\thanks{E-mail: ruffo@avanzi.de.unifi.it} }
\affiliation{1. Laboratoire de Physique, UMR-CNRS 5672, ENS Lyon, 46
  All\'{e}e d'Italie, 69364 Lyon c\'{e}dex 07, France\\
  2. Department of Physics, Tbilisi State University, 3 Chavchavadze
  avenue, Tbilisi 128, Republic of Georgia\\
  3.  Laboratoire de Biophysique Statistique, ITP, {\'E}cole Polytechnique
  F{\'e}d{\'e}rale de Lausanne, BSP CH-1015 Lausanne, Switzerland \\
  4.  Dipartimento di Energetica, ``S. Stecco'' and CSDC, Universit{\`a}
  di  Firenze, and INFN, via S. Marta, 3, 50139 Firenze, Italy}
\date{\today}

\begin{abstract}
  We present a detailed analysis of the modulational instability of
  the zone-boundary mode for one and higher-dimensional
  Fermi-Pasta-Ulam (FPU) lattices. Following this instability, a
  process of relaxation to equipartition takes place, which we have
  called the {\em Anti-FPU problem} because the energy is initially
  fed into the highest frequency part of the spectrum, at variance
  with the original FPU problem (low frequency excitations of the
  lattice).  This process leads to the formation of {\em chaotic
    breathers} in both one and two dimensions. Finally, the system
  relaxes to energy equipartition on time scales which increase as the
  energy density is decreased. We show that breathers formed when
  cooling the lattice at the edges, starting from a random initial
  state, bear strong qualitative similarities with chaotic breathers.
\end{abstract}

 \maketitle

{\bf Several nonlinear physical systems exhibit modulational
  instability, which is a self-induced modulation of the steady state
  resulting from a balance between nonlinear and dispersive effects.
  This phenomenon has been studied in a large variety of physical
  contexts: fluid dynamics, nonlinear optics and plasma physics. The
  Fermi-Pasta-Ulam (FPU) lattice is an extremely well--suited model
  system to study this process. Both the triggering of the instability
  and its further evolution can be studied in detail, exciting
  initially high-frequency modes. The original FPU problem was casted
  instead in the context of long wavelengths. This is why we call the
  process we analyze in this paper, the {\em Anti--FPU} problem
  because of the analogy with the seminal FPU numerical simulation. At
  variance with the appearance of (m)KdV-solitons in the FPU original
  problem, in this process the pathway to equipartition leads to the
  creation of localized objects that are {\em chaotic breathers}.
  Similar localized structures emerge when cooling the lattice at the
  edges, starting from thermalized initial states. }

\section{Introduction}

In 1955, reporting about one of the first numerical simulations,
Fermi, Pasta and Ulam (FPU)~\cite{FPU} remarked that it was {\em $\ldots$
  very hard to observe the rate of ``thermalization'' or mixing $\ldots$}
in a nonlinear one-dimensional lattice in which the energy was
initially fed into the lowest frequency mode.  Even if the
understanding of this problem advanced significantly
afterwards~\cite{ford,lichtlieb}, several issues are still far from
being clarified. In most cases, the evolution towards energy {\em
  equipartition} among linear modes has been checked considering an
initial condition where all the energy of the system is concentrated
in a small packet of modes centered around some low frequency.

Beginning with the pioneering paper of Zabusky and Deem~\cite{ZabuskyDeem},
the opposite case in which the energy is put into a high frequency
mode has been also analyzed. In this early paper, the zone--boundary mode
was excited with an added spatial modulation for the one-dimensional
$\alpha$-FPU model (quadratic nonlinearity in the equations of motion).
Here, we will study the time-evolution of this mode without any
spatial modulation for the $\beta$-FPU model (cubic nonlinearity in the
equations of motion) and some higher--order nonlinearities. Moreover,
we will extend the study to higher dimensional lattices. Since the energy
is fed into the opposite side of the linear spectrum, we call this problem
the {\em Anti-FPU problem}.

In a paper by Bundinsky and Bountis~\cite{bountis}, the
zone--boundary mode solution of the one-dimensional FPU lattice
was found to be unstable above an energy threshold $E_c$ which
scales like $1/N$, where $N$ is the number of oscillators. This result
was later and independently confirmed by Flach~\cite{flach} and Poggi
et al.~\cite{PR}, who also obtained the correct factor in the large
$N$-limit. These results were obtained by a direct linear stability
analysis around the periodic orbit corresponding to the zone-boundary
mode. Similar methods have been recently applied to other modes and
other FPU-like potentials by Chechin et al~\cite{chechin,chechin2004}
and Rink~\cite{Rink2003}.

A formula for $E_c$, valid for all $N$, has been obtained in
Refs.~\cite{sandusky,burlakov,DRT,Paladin} in the rotating wave
approximation, and will be also discussed  in this paper.  Associated
with this instability is the calculation of the growth rates of mode
amplitudes. The appropriate approach for Klein-Gordon lattices was first
introduced by Kivshar and Peyrard~\cite{KivsharPeyrard}, following an
analogy with the Benjamin-Feir instability in fluid
mechanics~\cite{benjaminFeir}.

Previously, a completly different approach to describe this
instability was introduced by Zakharov and Shabat~\cite{zakharov},
studying the associated Nonlinear Schr{\"o}dinger equation in the
continuum limit. A value for the energy threshold was obtained in
Ref.\cite{lukomskii} in the continuum limit. The full derivation
starting from the FPU equation of motions was then independently obtained
by Berman and Kolovskii~\cite{berman} in the so-called
``narrow-packet'' approximation.

Only very recently the study of what happens after 
modulational instability develops has been performed for
Klein-Gordon~\cite{daumont} and
FPU-lattices\cite{maledetirusi,sandusky}. From these analyses it
turned out that these high-frequency initial conditions lead to a
completely new dynamical behavior in the transient time preceeding
the final energy equipartition.  In particular, the main discovery
has been the presence on the lattice of sharp localized
modes~\cite{maledetirusi,daumont}.  These latter papers were the
first  to make the connection between energy relaxation and
intrinsic localized modes~\cite{ST}, or
breathers~\cite{diffflach1,diffflach2,diffflach3,reviewbreather}.
Later on, a careful numerical and theoretical study of the
dynamics of a $\beta$-FPU model was performed~\cite{cretegny}. It
has been shown that moving breathers play a relevant role in the
transient dynamics and that, contrary to exact breathers, which
are periodic solutions, these have a chaotic evolution. This is
why they have been called {\em chaotic
  breathers}. Following these studies, Lepri and
Kosevich~\cite{kosevichlepri} and Lichtenberg and
coworkers~\cite{ulman,mirnov} have further characterized the scaling
laws of relaxation times using also continuum limit equations.

On the other hand, studies of the asymptotic state of the FPU lattice
dynamics when energy is extracted from the boundaries have revealed
the persistence of localized
modes~\cite{Aubry1,Aubry2,noi,reigada1,noichaos}. Already some of
these authors~\cite{noi,noichaos} have discussed the similarities of
these modes with chaotic breathers. In this paper, we will further
study this connection.

Most of the previous studies are for one-dimensional lattices. Here, we
will derive modulational instability thresholds also for higher
dimensional lattices and we will report on a study of chaotic
breathers formation in two-dimensional FPU lattices.

We have organized the paper in the following way.  In
Section~\ref{modinst}, the modulational instability of zone-boundary
modes on the lattice is discussed, beginning with the one-dimensional
case, followed by the two-dimensional and higher dimensional cases and
finishing with the continuum Nonlinear Schr{\"o}dinger approach.
Section~\ref{localization1d} deals with the mechanisms of creation of
chaotic breathers in  one  and two dimensions.  Finally,
in Section~\ref{Spontaneouslocalization}, we discuss the relation with
numerical experiments performed when the lattice is cooled at the
edges. Some final remarks and conclusions are reported in
Section~\ref{conclusions}.

\section{Modulational Instability}
\label{modinst}

\subsection{The one-dimensional case}
\label{modinst1D}

We will discuss in this section modulational instability for the
one-dimensional FPU lattice, where the linear coupling is corrected by
a $(2p+1)$-th order nonlinearity, with $p$ a positive integer.  Denoting
by $u_n(t)$ the relative displacement of the $n$-th particle from its
equilibrium position, the equations of motion are
\begin{eqnarray}
\ddot{u}_n = u_{n+1} + u_{n-1} - 2u_n +
(u_{n+1}-u_n)^{2p+1} - (u_n - u_{n-1})^{2p+1}.
\label{sub}
\end{eqnarray}
We adopt a lattice of $N$ particles and we choose periodic boundary
conditions.  For the sake of simplicity, we first report on  the analysis for
$p=1$ (i.e. the  $\beta$-FPU model) and then we generalize the
results to any $p$-value.

Due to periodic boundary conditions, the normal modes
associated to the linear part of Eq.~(\ref{sub})
are plane waves of the form
\begin{equation}
u_n(t) =\frac{a}{2}\left(e^{i\theta_n (t)}+e^{-i\theta_n (t)}\right)
\label{plane}
\end{equation}
where $\theta_n(t) = qn-\omega t$ and $q=2\pi k/N$ ($k=-N/2,\dots ,N/2$).  The
dispersion relation of nonlinear phonons in the rotating wave
approximation~\cite{sandusky} is $\omega^2(q)=4(1+\alpha)\sin ^{2} (q/2)$, where $\alpha=3a^2
\sin^2(q/2)$ takes into account the nonlinearity.  Modulational
instability of such a plane wave is investigated by studying the
linearized equation associated with the envelope of the carrier
wave~(\ref{plane}).  Therefore, one introduces infinitesimal
perturbations in the amplitude and phase and looks for solutions of
the form
\begin{eqnarray}
u_n(t) &=& \frac{a}{2}[ 1 + b_n (t) ]\ e^{i\left[\theta_n (t)+\psi_n(t)\right]}+
 \frac{a}{2}[ 1 + b_n (t) ]\ e^{-i\left[\theta_n (t)+\psi_n(t)\right]}\nonumber\\
&=&a[1+b_n(t)]\ \cos[qn-\omega t+\psi_n(t)],
\label{az}
\end{eqnarray}
where  $b_n$ and $\psi_n$ are reals and assumed to be small
in comparison with the parameters of the carrier wave.
Substituting Eq.~(\ref{az}) into the equations of motion and
keeping the second derivative,
we obtain for the real and imaginary part of the secular term
$e^{i(qn-\omega t)}$ the following equations
\begin{eqnarray}
-\omega^2b_n+2\omega\dot\psi_n+\ddot b_n&=&(1+2\alpha)
\left[\cos q\,(b_{n+1}+ b_{n-1})-2b_n \right]\nonumber \\
&-&\alpha\left(b_{n+1}+ b_{n-1}-2b_n\cos q \right)
-(1+2\alpha)\sin q\,(\psi_{n+1}-\psi_{n-1}) \label{eq4} \\
-\omega^2\psi_n-2\omega\dot b_n+\ddot \psi_n&=&
(1+2\alpha)\left[\cos q\, (\psi_{n+1}+\psi_{n-1})-2\psi_n\right]
\nonumber \\
&+&(1+2\alpha)\sin q\,(b_{n+1}-b_{n-1})+\alpha
\left(\psi_{n+1}+ \psi_{n-1}-2\psi_n\cos q  \right).\label{eq5}
\end{eqnarray}

Further assuming  $b_n=b_0 \ e^{i(Qn-\Omega t)}+{\rm c.c.}$
and $\psi_n=\psi_0 \ e^{i(Qn-\Omega t)}+{\rm c.c.}$
we obtain  the two following equations for the secular term
$e^{i(Qn-\Omega t)}$
\begin{eqnarray}
b_0\Bigl[\Omega^2+\omega^2+2(1+2\alpha)(\cos q \cos Q -1)
-2\alpha(\cos Q-\cos q) \Bigr]
-2i\psi_0\left[\omega\Omega+(1+2\alpha)\sin q\sin Q \right]=0\label{eq6}\\
\psi_0\Bigl[ \Omega^2+\omega^2+2(1+2\alpha)(\cos q\cos Q-1)
+2\alpha(\cos Q-\cos q)\Bigr]
+2ib_0\left[ \omega\Omega +(1+2\alpha)\sin q\sin Q\right]=0.\label{eq7}
\end{eqnarray}
In the case of Klein-Gordon type
equations~\cite{KivsharPeyrard,daumont}, one neglects the second order
derivatives in Eqs.~(\ref{eq4})-(\ref{eq5}). This can be justified by
the existence of a gap in the dispersion relation for $q=0$, which
allows to neglect $\Omega^2$ with respect to $\omega^2$. In the FPU case, this
approximation is worse, especially for long wavelengths, because there
is no gap.

Non trivial solutions of Eqs.~(\ref{eq6})-(\ref{eq7}) can be
found only if the Cramer's determinant vanishes, i.e. if the
following equation is fulfilled:
\begin{eqnarray}
\Biggl[(\Omega+\omega)^2-4(1+2\alpha)\sin^2\left({q+Q\over2}\right)\Biggr]&&
\Biggl[(\Omega-\omega)^2-
4(1+2\alpha)\sin^2\left({q-Q\over2}\right)\Biggr]  \nonumber \\
&=& 4\alpha^2\left(\cos{Q}-\cos q\right)^2.
\label{relatdispercorr}
\end{eqnarray}

This equation admits four different solutions when the wavevectors $q$
of the unperturbed wave and $Q$ of the perturbation are fixed.  If one
of the solutions is complex, an instability of one of the modes $(q\pm
Q)$ is present, with a growth rate given by the imaginary part of the
solution. Using this method, one can derive the instability threshold
amplitude for any wavenumber. A trivial example is the $q=0$ case,
for which we obtain $\Omega=\pm \sin\left({Q/ 2}\right)$, which proves that
the zero mode solution is stable. This mode is present due to the
invariance of the equations of motion~(\ref{sub}) with respect to the
translation $u_n \to u_n+const$ and, as expected, is completely
decoupled from the others.

\begin{figure}[ht]
\resizebox{0.5\textwidth}{!}{\includegraphics{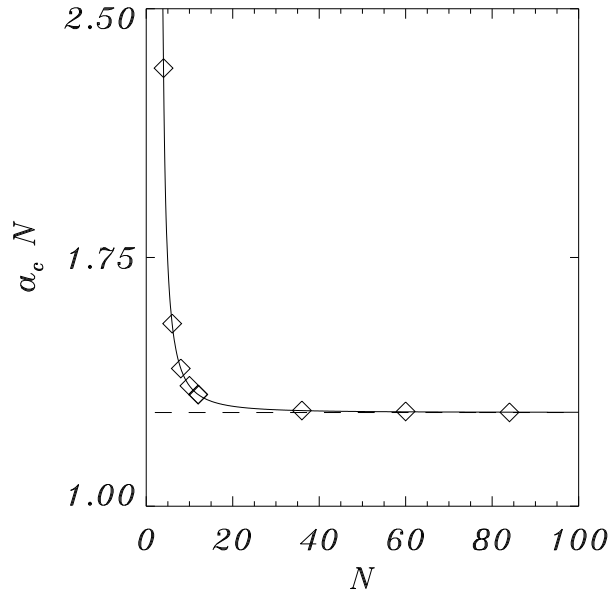}}
\caption{Modulational instability threshold amplitude for the $\pi$-mode
versus the number of particles in the one-dimensional FPU lattice. The solid line
  corresponds to the analytical formula~(\ref{final1d}), the dashed
  line to its large $N$-estimate~(\ref{final1dapp}) and the diamonds
  are obtained from numerical simulations.}
\label{modinstab1d}
\end{figure}

A first interesting case is $q=\pi$. One can easily see that
Eq.~(\ref{relatdispercorr}) admits two real and two complex conjugate
imaginary solutions if
and only if
\begin{equation}
\cos^2{Q\over 2}>{1+\alpha\over 1+3\alpha}.
\label{acrit}
\end{equation}
This formula was first obtained by Sandusky and Page (Eq.~(22) in
Ref.~\cite{sandusky}) using the rotating wave approximation.  The
first mode to become unstable when increasing the amplitude $a$
corresponds to the wavenumber $Q={2\pi/N}$.  Therefore, the critical
amplitude $a_c$ above which the $q=\pi$-mode looses stability is
\begin{equation}
a_c= \left( {\sin^2\left({\pi/ N}\right)\over 3
\left[3\cos^2\left({\pi/ N}\right)-1\right]} \right)^{1/2}\label{final1d}.
\end{equation}
This formula is valid for all even values of $N$ and its large
$N$-limit is
\begin{equation}
a_c= \frac{\pi}{\sqrt{6}N}+O\left({1\over {N^3}}\right)\label{final1dapp}.
\end{equation}
In Fig.~\ref{modinstab1d}, we show its extremely good agreement
with the critical amplitude determined from numerical simulations.
It is interesting to emphasize that the analytical
formula~(\ref{final1d}) diverges for $N=2$, predicting that the
$\pi$-mode is stable for all amplitudes in this smallest lattice.
This is in agreement with the Mathieu equation analysis  (see
Ref.~\cite{PR} p.~265).

It is also interesting to express this result in terms of the total
energy to compare with what has been obtained using other methods
\cite{zakharov,bountis,berman,flach,PR}. Since for the $\pi$-mode the
energy is given by $E=N(2a^2+4a^4)$, we obtain the critical
energy
\begin{equation}
E_c={2N\over 9}\sin^2\left({\pi\over
N}\right){7\cos^2\left({\pi/ N}\right)-1
\over \left[3\cos^2\left({\pi/ N}\right)-1\right]^2}.
\label{eqnc}
\end{equation}
For large $N$, we get
\begin{equation}
E_c={\pi^2\over 3N}+O\left({1\over {N^3}}\right).\label{scalingec}
\end{equation}
This asymptotic behavior is the same as the one obtained using the
narrow packet approximation in the context of the Nonlinear
Schr{\"o}dinger equation by Berman and Kolovskii (Eq.~(4.1) in
Ref.~\cite{berman}). The correct scaling behavior with $N$ of the
critical energy has been also obtained by Bundinsky and Bountis
(Eq.~(2.22) in Ref.~\cite{bountis}) by a direct linear stability
analysis of the $\pi$-mode. The correct formula, using  this latter method,
has been independently obtained by Flach (Eq.~(3.20)) in
Ref.~\cite{flach}) and Poggi and Ruffo (p. 267 of Ref.~\cite{PR}).
Recently, the $N^{-1}$-scaling of formula~(\ref{scalingec}) has been confirmed
using a different numerical method and, interestingly, it holds also for
the  $2\pi/3$ and $\pi/2$ modes~\cite{Cafarella}.

This critical energy is also very close to the Chirikov
``stochasticity threshold'' energy obtained by the resonance overlap
criterion for the zone boundary mode\cite{izraeilchirikov}. The
stochasticity threshold phenomenon has been thoroughly studied for
long wavelength initial conditions, and it has been clarified that it
corresponds to a change in the scaling law of the largest Lyapunov
exponent\cite{pettini}.  We will show in Section \ref{localization1d}
that above the modulational instability critical energy for the
$\pi$-mode one reaches asymptotically a chaotic state with a positive Lyapunov
exponent, consistently with Chirikov's result.

The above results can be generalized to nonlinearities of $2p+1$ order
in the equations of motion~(\ref{sub}). We limit the analysis to the $\pi$-mode,
for which the instability condition~(\ref{acrit}) takes the form
\begin{equation}
\cos^2{Q\over 2}>{1+\alpha\over 1+(2p+1)\alpha},
\label{acritnew}
\end{equation}
where
\begin{equation}
\alpha=\frac{(2p+1)!}{p ! (p+1) !}\, a^{2p}.
\end{equation}
Hence the critical amplitude above which the
$\pi$-mode is unstable is
\begin{equation}
a_c= \left[ {p ! (p+1) !\, \sin^2\left({\pi/ N}\right)\over (2p +1) !
\left[(2p +1)\cos^2\left({\pi/ N}\right)-1\right]} \right]^{1/2}\label{final1dpourp},
\end{equation}
leading to the large $N$ scaling
\begin{eqnarray}
a_c&\sim& N^{-{1/ p}}\\
E_c&\sim& N^{1-2/p}.
\end{eqnarray}
This scaling also corresponds to the one found in Ref.~\cite{Kladko}
when discussing tangent bifurcations of band edge plane waves in
relation with energy thresholds for discrete breathers. Their
``detuning exponent'' $z$ has a direct connection with the nonlinearity
exponent $p=z/2$. We will see in Section~\ref{modinst2D} that this
analogy extends also to higher dimensions.

For fixed $N$, $a_c$ is an increasing function of the power of the
coupling potential with the asymptotic limit $\lim_{p\to\infty}
a_c=0.5$. Therefore, in the hard potential limit the critical
energy for the $\pi$-mode increases proportionally to $N$. The
fact that we find a higher energy region where the system is
chaotic is not in contradiction with the integrability of the
one-dimensional system of hard rods\cite{refhardrod}, because in
the present case we have also a harmonic contribution at small
distances.

For the FPU-$\alpha$ model (quadratic nonlinearity in the equations of
motion), the $\pi$-mode is also an exact solution which becomes unstable
at some critical amplitude which, contrary to the case of the FPU-$\beta$
model, is $N$-independent\cite{sandusky,chechin}; which means that the
critical energy is proportional to $N$ and then that $\pi$-mode can be
stable in some low energy density limit also in the thermodynamic
limit.

It has also been
realized~\cite{PR,marie,julien,chechin,Shinohara,Rink2003} that group
of modes form sets which are invariant under the dynamics.  The
stability analysis~\cite{julien,chechin2004} of pair of modes has
shown a complex dependence on their relative amplitudes.  The
existence of such invariant manifolds has also allowed to construct
Birkhoff-Gustavson normal forms for the FPU model, paving the way to
KAM theory~\cite{Rink2000}.

\subsection{Higher dimensions}
\label{modinst2D}

In this section, we will first discuss modulational instability of the
two-dimensional FPU model. The method presented in
section~\ref{modinst1D} can be easily extended and the global physical
scenario is preserved. However, the scaling with $N$ of the critical
amplitude changes in such a way to make  critical energy constant, in
agreement with the analysis of Ref.~\cite{Kladko}.

The masses lie on a two-dimensional square lattice with unitary
spacing in the $(x,y)$ plane. We consider a small relative
displacement $u_{n,m}$ ($n,m\in [1,N]$) in the vertical $z$-direction.
Already with a harmonic potential, if the spring length at
equilibrium is not unitary, the series expansion in $u_{n,m}$ of the
potential contains all even powers. We retain only the first two terms
of this series expansion. After an appropriate rescaling of time and
displacements to eliminate mass and spring constant values, one gets
the following adimensionallized equations of motions
\begin{eqnarray}
\ddot u_{n,m}&=&u_{n+1,m}+u_{n-1,m}+u_{n,m+1}+u_{n,m-1}-4u_{n,m} \nonumber \\ &&
+\left(u_{n+1,m}- u_{n,m}\right)^3+ \left(u_{n-1,m}-u_{n,m}\right)^3 +\left(u_{n,m+1}- u_{n,m}\right)^3+\left(u_{n,m-1}- u_{n,m}\right)^3 .\label{2Dfpu}
\end{eqnarray}
Considering periodic boundary conditions, plane waves solutions have
the form
\begin{equation}
u_{n,m}=a\cos\left(q_xn+q_ym-\omega t\right). \label{harmonic}
\end{equation}
In the rotating wave approximation\cite{sandusky}, one immediately
obtains the dispersion relation
\begin{eqnarray}
\omega^2=4\sin^2\frac{q_x}{2}+4\sin^2\frac{q_y}{2}+12a^2\left[\sin^4
\frac{q_x}{2}+\sin^4\frac{q_y}{2}\right],
\label{dispersion}
\end{eqnarray}
which becomes exact for the zone-boundary mode $(q_x, q_y)=(\pi,\pi)$,
\begin{equation}
\omega_{\pi,\pi}^2=8(1+3a^2).
\label{2ddisp}
\end{equation}
In order to study the stability of the zone-boundary mode, we adopt a
slightly different approach. Namely, we consider the perturbed
relative displacement field of the form
\begin{equation}
u_{n,m}=\left(\frac{a}{2}+b_{n,m}\right)e^{i(\pi n+\pi m-\omega_{\pi,\pi}t)}+
c.c.,
\label{disp}
\end{equation}
where $b_{n,m}$ is complex. This approach turns out to be equivalent
to the one of Section~\ref{modinst1D} in the linear limit.

\begin{figure}[t]
\resizebox{0.5\textwidth}{!}{\includegraphics{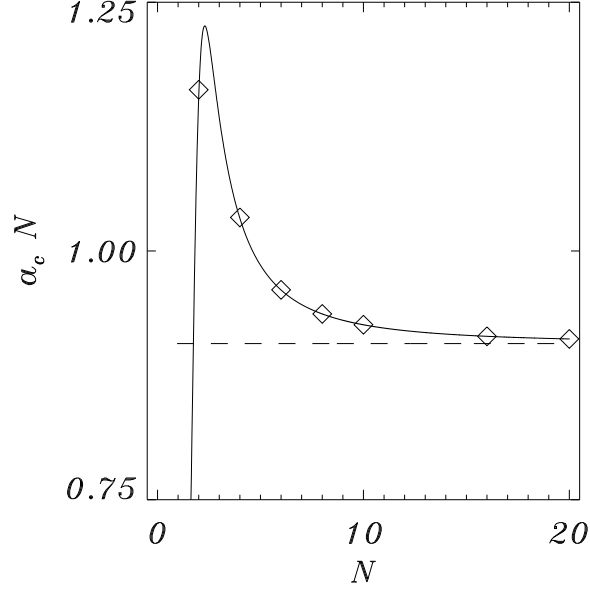}}

\caption{Modulation instability threshold for the $(\pi,\pi)$ mode versus number
  of oscillators in two dimensional array. The solid line is given by
  the corresponds to the estimate obtained from the Nonlinear
  Schr{\"o}dinger equation in large $N$ limit (see formula (\ref{form}))
  and the diamonds are results of numerical simulations.}
\label{modinstab2d}
\end{figure}

Substituting this perturbed displacement field in Eqs.~(\ref{2Dfpu}), we obtain
\begin{eqnarray}
[1+2\alpha]\left[ b_{n+1,m}+ b_{n-1,m}+ b_{n,m+1}+ b_{n,m-1}+ 4b_{n,m}\right] &&
\nonumber \\
-\alpha\left[b^*_{n+1,m}+ b^*_{n-1,m}+ b^*_{n,m+1}+ b^*_{n,m-1}+
4b^*_{n,m}\right]&&
=-\ddot b_{n,m}+2i\omega_{\pi,\pi}\dot b_{n,m}+\omega_{\pi,\pi}^2b_{n,m},
\label{bnm}
\end{eqnarray}
where $\alpha=3a^2$. Looking for  solutions of the form
\begin{equation}
b_{n,m}=Ae^{i(Q_xn+Q_ym-\Omega t)}+ Be^{-i(Q_xn+Q_ym-\Omega t)},
\end{equation}
we arrive at the following set of linear algebraic equations for the complex
constants $A$ and $B$
\begin{eqnarray}
\left[(\Omega+\omega_{\pi,\pi})^2-
8(1+2\alpha)\Delta\right]A+8\alpha \Delta B&=&0  \\
8\alpha \Delta A+ \left[(\Omega-\omega_{\pi,\pi})^2-
8(1+2\alpha)\Delta\right]B&=&0,
\label{eq}
\end{eqnarray}
where $2\Delta=\cos^2(Q_x/2)+\cos^2(Q_y/2)$. As for the
one-dimensional case, we require that the determinant of this linear
system in $A$ and $B$ vanishes, which leads to the following condition
\begin{eqnarray}
\left[(\Omega+\omega_{\pi,\pi})^2-8\Delta (1+2\alpha)\right]
\left[(\Omega-\omega_{\pi,\pi})^2-
8\Delta(1+2\alpha)\right] = 64\alpha^2\Delta^2.
\label{relatdispercorr2D}
\end{eqnarray}
This equation admits two real and two complex conjugated
imaginary solutions in $\Omega$ if
\begin{equation}
\Delta>\frac{1+\alpha }{1+3\alpha },
\end{equation}
which is the analogous for two dimensions of condition (\ref{acrit}).
One can achieve the minimal nonzero value of the r.h.s. of the above expression choosing
$Q_x=0$, $Q_y=2\pi/{N}$, which leads to the following result for the
critical amplitude
\begin{equation}
a_c=\left(\frac{\sin^2(\pi/{N})}{3[3\cos^2(\pi/{N})+1]}\right)^{1/2}.
\label{final}
\end{equation}
Its large $N$ limit is
\begin{equation}
a_c= \frac{\pi}{\sqrt{12}N}+O\left({1\over {N^3}}\right)\label{final2dapp}.
\end{equation}
This prediction is compared with numerical data in
Fig.~\ref{modinstab2d}.
The agreement is good for all values of $N$.

Since the relation between energy and amplitude is now
$E=2N^2(2a^2+4a^4)$,
we obtain the critical energy in the large $N$-limit as
\begin{equation}
E_c={\pi^2\over 3}+O\left({1\over {N^2}}\right).\label{scalingec2D}
\end{equation}
This shows that the critical energy is now constant in the
thermodynamic limit, which agrees with the remark of Ref.~\cite{Kladko}
about the existence of a minimal energy for breathers formation~\cite{reviewbreather}.

The results of this Section can be easily extended to any dimension
$d$.  Repeating the same argument, we arrive at the following estimates
for the critical amplitude and energy in the large $N$ limit
\begin{eqnarray}
a_c&=& \frac{\pi}{\sqrt{6d}}\frac{1}{N}+O\left({1\over {N^3}}\right)\label{acinfini}\\
E_c&=&{\pi^2\over 3}N^{d-2}+O\left({N^{d-4}}\right).\label{ecinfini}
\end{eqnarray}
This means that the critical energy density $ \varepsilon_c=E_c/N$ for destabilizing the
zone boundary mode vanishes as $1/N^2$, independently of dimension.

\subsection{Large $N$ limit using the Nonlinear Schr{\"o}dinger equation}

The large $N$ limit expressions~(\ref{acinfini}) and~(\ref{ecinfini})
can be derived also by continuum limit considerations. We will derive
the general expression for any dimension $d$. The
displacement field can be factorized into a complex envelope part $\psi$
multiplied by the zone boundary mode pattern in $d$ dimensions.
\begin{equation}
u_{n_1,\dots,n_d }=\frac{\psi(n_1,\dots,n_d,t)}{2}e^{i\left(\pi \sum_{ i=1}^dn_i-\omega_{\pi,
    \dots , \pi }t\right)}+ c.c.\ ,
\label{dispp}
\end{equation}
where
\begin{equation}
\omega_{\pi,\dots , \pi }=\sqrt{d\left[4(1+3|\psi|^2)\right]}.
\label{dispo}
\end{equation}
Substituting Eq.~(\ref{dispp}) into the FPU lattice equations in $d$ dimensions,
a standard procedure~\cite{RemoissenetSemiDis,book} leads to the following
$d$ dimensional Nonlinear Schr{\"o}dinger (NLS) equation:
\begin{equation}
i \frac{\partial \psi }{\partial t} + \frac{P}{2}\,
\Delta_d \psi- Q \psi
{\left| \psi \right|}^2 =0,
\label{nls}
\end{equation}
where $\Delta_d$ is the $d$ dimensional Laplacian. The parameters $P$ and
$Q$ are derived from the nonlinear dispersion relation
\begin{eqnarray}
\omega^2=\sum_{i=1}^{d}\left[4\sin^2\frac{q_{i}}{2}+12|\psi|^2\sin^4
\frac{q_{i}}{2}\right],
\label{dispersionddim}
\end{eqnarray}
as
\begin{eqnarray}
P&=&\frac{\partial^2\omega}{\partial
q_i^2}\left({q_1=\pi,\dots, q_d=\pi,|\psi| =0}\right)=\frac{1}{2\sqrt{d}}  \\
Q&=&-\frac{\partial\omega}{\partial
|\psi|^2}\left({q_1=\pi,\dots, q_d=\pi,|\psi| =0}\right)=-3\sqrt{d}.
\label{para}
\end{eqnarray}

Assuming that, at the first stage, modulational  instability develops
along a single direction $x$ and that the field remains constant along all
other directions, one gets  the one-dimensional NLS equation
\begin{equation}
i \frac{\partial \psi }{\partial t} +\frac{P}{2}
\frac{\partial^2 \psi }{\partial x^2}- Q \psi
{\left| \psi  \right|}^2 =0.
\label{nls1}
\end{equation}
Following the results of the inverse scattering approach
\cite{zakharov}, any initial distribution of amplitude $|\psi|$ and
length $\lambda $ along $x$, and constant along all other directions,
produces a final localized distribution if \cite{lukomskii}
\begin{equation}
(|\psi|\lambda)^2>\pi^2\left|\frac{P}{Q}\right|.
\label{form0}
\end{equation}
This means that if the initial state is taken with constant amplitude
$|\psi|=a$ on the $d$-dimensional lattice with $N^d$ oscillators, the
modulational instability threshold is
\begin{equation}
(a_c{N})^2=\frac{\pi^2}{6d}
\label{form}
\end{equation}
which coincides with the leading order in Eq. (\ref{acinfini}).

\section{Chaotic Breathers}
\label{localization1d}

In this Section, we will discuss what happens when the
modulational energy threshold is overcome. The first thorough
study of this problem can be found in Ref.~\cite{maledetirusi},
many years after the early pioneering work of Zabusky and
Deem~\cite{ZabuskyDeem}. Already in Ref.~\cite{maledetirusi}, it
has been remarked that an energy localization process takes place,
which leads to the formation of breathers~\cite{reviewbreather}.
This process has been further characterized in terms of
time-scales to reach energy equipartition and of quantitative
localization properties in Ref.~\cite{cretegny}. The localized
structure which emerges after modulational instability has been
here called ``chaotic breather'' (CB).  The connection between CB
formation and continuum equations has been discussed in
Refs.~\cite{kosevichlepri,mirnov}, while the relation with the
process of relaxation to energy equipartition has been further
studied in Ref.~\cite{ulman}.  We will briefly recall some
features of the localization process in one dimension and present
new results for two dimensions.

For long time simulations, we use appropriate symplectic
integration schemes in order to preserve as far as possible the
Hamiltonian structure. For the one dimensional FPU, we adopt a
6th-order Yoshida's algorithm~\cite{yoshida} with a time step $dt
= 0.01$; this choice allows us to obtain an energy conservation
with a relative accuracy $\Delta E / E$ ranging from $10^{-10}$ to
$10^{-12}$. For two dimensions, we use instead the 5--th order
symplectic Runge--Kutta--Nystr\"{o}m algorithm of
Ref.~\cite{Calvo}, which gives a similar quality of energy
conservation.

\begin{figure}
\resizebox{0.5\textwidth}{!}{\includegraphics{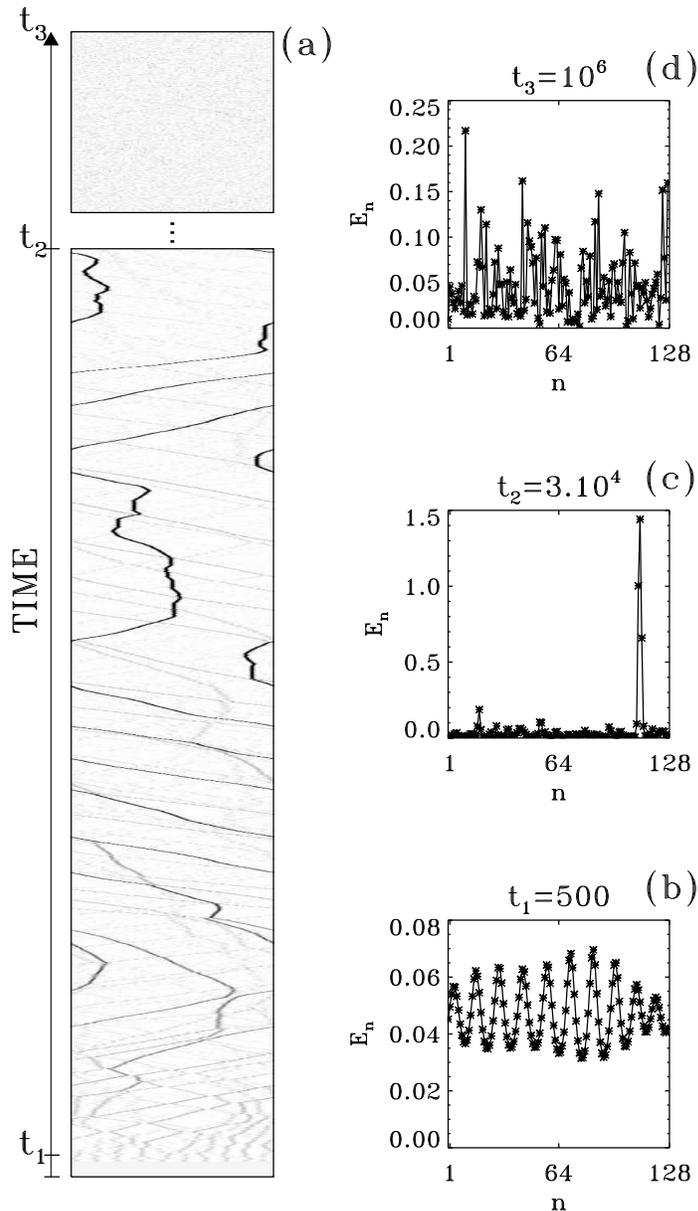}}
\caption{Time evolution of the local energy (\ref{localenergy}).
  In panel~(a), the horizontal axis indicates lattice sites and the
  vertical axis is time. The grey scale goes from $E_n=0$ (white) to
  the maximum $E_n$-value (black). The lower rectangle corresponds to
  $0<t<3\ 10^4$ and the upper one to $0.994\ 10^6<t<10^6$. Figs.~(b),
  (c) and~(d) show the instantaneous local energy $E_n$ along the
  $N=128$ chain at three different times. Remark the difference in
  vertical amplitude in panel~(c), when the chaotic breather is
  present.  The initial $\pi$-mode amplitude is $a=
  0.15>a_c\simeq0.01$.}
\label{greyscales}
\end{figure}

We report in Fig.~\ref{greyscales}(a) a generic evolution of the
one dimensional $\pi$-mode above the  modulation instability
critical amplitude ($ a > a_c$). The grey scale refers to the
energy residing on site~$n$,
\begin{equation}
E_n = {1\over 2} \dot{u}_n^2 + {1\over 2}V(u_{n+1}-u_n) +
{1\over 2} V(u_n-u_{n-1}),\label{localenergy}
\end{equation}
where the FPU-potential is $V(x) = {1\over 2}x^2 + {1\over 4} x^4$.
Figs.~\ref{greyscales}(b), \ref{greyscales}(c) and \ref{greyscales}(d)
are three successive snapshots of the local energy $E_n$ along the
chain. At short time, a slight modulation of the energy in the system
appears (see Fig.~\ref{greyscales}(b)) and the $\pi$-mode is
destabilized~\cite{sandusky}. Later on, as shown in
Fig.~\ref{greyscales}(a), only a few localized energy packets emerge:
they are breathers~\cite{reviewbreather}.  Inelastic collisions of
breathers have a systematic tendency to favour the growth of the big
breathers at the expense of small ones~\cite{prllocali,BangPeyrard}.
Hence, in the course of time, the breather number decreases and only
one, of very large amplitude, survives (see Fig.~\ref{greyscales}(c)):
this is the localized excitation we have called chaotic breather (CB).
The CB moves along the lattice with an almost ballistic motion:
sometimes it stops or reflects. During its motion the CB collects
energy and its amplitude increases. It is important to note that the
CB is never at rest and that it propagates with a given subsonic
speed~\cite{KosevichCorso}.  Finally, the CB decays and the system
reaches energy equipartition, as illustrated in
Fig.~\ref{greyscales}(d). The final CB decay is present in all the
simulations we have performed, but one cannot exclude the existence of
examples where the final breather does not disappear.

\begin{figure}
\resizebox{0.75\textwidth}{!}{\includegraphics{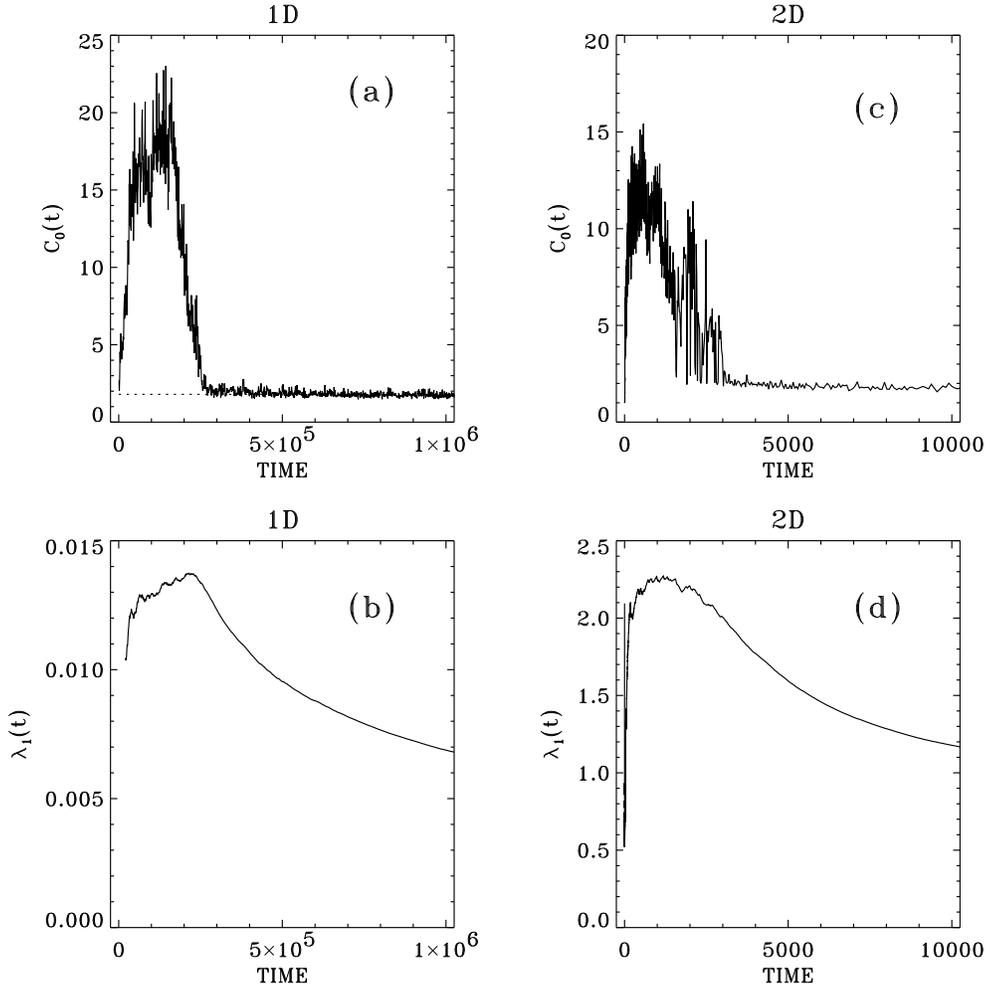}}

\caption{Panel~(a) presents the evolution of $C_0(t)$ of formula
  (\ref{clocal}) for the one-dimensional FPU lattice with $N=128$
  oscillators, initialized on the $\pi$-mode with an amplitude
  $a=0.126>a_c\simeq0.010$. The dashed line indicates the equilibrium value
  $\bar C_0=1.795$.  Panel~(b) presents the corresponding finite time
  largest Lyapunov exponent. Panel~(c) shows $C_0(t)$ for the
  two-dimensional FPU lattice with $20*20$ oscillators, initialized on
  the $(\pi,\pi)$-mode with an amplitude $a=0.425>a_c\simeq0.045$.
  Panel~(d) presents the finite time largest Lyapunov exponent for two
  dimensions.}
\label{cohlyap}
\end{figure}

In order to obtain a quantitative characterization of
energy localization, we introduce the ``participation ratio''
\begin{equation}
C_0(t)=N {\displaystyle \sum_{i=1}^N E_i^2\over
\left(\displaystyle \sum_{i=1}^N E_i  \right)^2 },
\label{clocal}
\end{equation}
which is of order one if $E_i = E/N$ at each site of the chain and of
order $N$ if the energy is localized on only one site. In
Fig.~\ref{cohlyap}(a), $C_0$ is reported as a function of time.
Initially, $C_0$ grows, indicating that the energy, evenly distributed
on the lattice at $t=0$, localizes over a few sites.  This localized
state survives for some time. At later times, $C_0$ starts to decrease
and finally reaches an asymptotic value $ \bar C_0$ which is
associated with the disappearance of the CB (an estimate of $ \bar
C_0$ has been derived in Ref.~\cite{cretegny} taking into account
energy fluctuations and is reported with a dashed line in
Fig.~\ref{cohlyap}(a)).  At this stage, the energy distribution in
Fourier space is flat, i.e.  a state of energy equipartition is
reached.

As explained in Ref.~\cite{cretegny}, the destruction of the breathers
is related to its interaction with low frequency modes which are
dominant in the chain after the initial stage. A full study of the
scattering of plane waves by FPU-breathers would be necessary to
quantify this explanation either as anticipated in
Ref.~\cite{cretegny} or, even better, as recently performed by Flach
and collaborators for the discrete NLS equation, Klein-Gordon or FPU
chains~\cite{flachandrei,flachandreiFPU}.

In Fig.~\ref{cohlyap}(b), we show the finite time largest Lyapunov
exponent $\lambda_1(t)$ for the same orbit as in Fig.~\ref{cohlyap}(a). We
observe a growth of $\lambda_1(t)$ when the CB emerges on the lattice and a
decrease when it begins to dissolve. The peak in $\lambda_1(t)$ perfectly
coincides with the one in $C_0$. Due to this increase of chaos
associated with localization, we have called the breather chaotic
(although chaos increase could be the result of a more complicated
process of interaction with the background).

In Ref.~\cite{cretegny}, the time-scale for the relaxation to
equipartition has been found to increase as $(E/N)^{-2}$ in the small
energy limit. This has been confirmed by the followers of this
study~\cite{kosevichlepri,ulman,mirnov}. Such power law scalings are
found also for the FPU relaxation starting from long
wavelengths~\cite{delucca}: the so-called {\it FPU problem}.  We have
termed the relaxation process which starts from short wavelengths the
{\it Anti-FPU problem}, just because of the similarities in the
scaling laws. The main feature of the latter problem is that
relaxation to equipartition goes through a complex process of
localized structures formation well described by breathers or, in the
low-amplitude limit, by solitons of the NonLinear Schrodinger
equation. On the contrary, for the original FPU problem, an initial
long wavelength excitation breaks up into a train of mKdV-solitons.
The final relaxation to equipartition is however due to an energy
diffusion process which has similar features for both the FPU and the
Anti-FPU problem~\cite{ulman}.

\begin{figure}[ht]
\resizebox{0.75\textwidth}{!}{\includegraphics{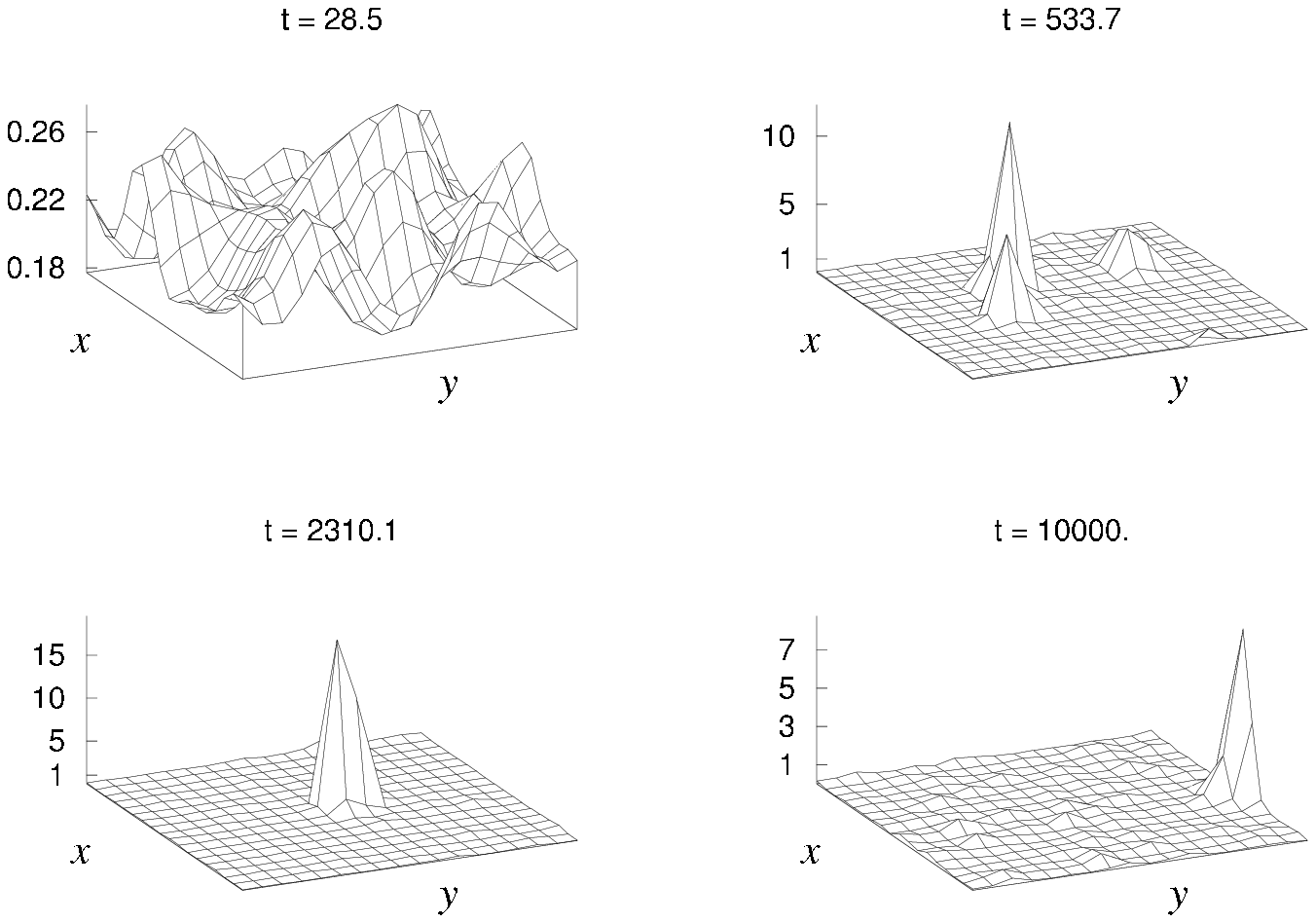}}

\caption{Local energy (\ref{localenergy2d})  surface plots  for the
  two-dimensional FPU lattice with $20*20$ oscillators, initialized on
  the $(\pi,\pi)$-mode with an amplitude $a=0.225>a_c\simeq0.045$.
  Snapshots at four different times $t$ are shown. Breathers form
  after a coalescence process similarly to the one-dimensional
  case. The mobility of the breathers is evident and one also observes
  in the last panel the final decrease.}
\label{fran1}
\end{figure}

A similar evolution of the local energy
\begin{eqnarray}
E_{n,m} = {1\over 2} \dot{u}_{n,m}^2 &&+ {1\over 4}V(u_{n+1,m}-u_{n,m})
+ {1\over 4}V(u_{n,m+1}-u_{n,m}) \nonumber\\
&&+ {1\over 4}V(u_{n-1,m}-u_{n,m}) + {1\over 4}V(u_{n,m-1}-u_{n,m}) \label{localenergy2d}
\end{eqnarray}
is observed for the two-dimensional case (see Fig. \ref{fran1}).
In this figure, we just show the initial evolution which leads to
breathers formation. As for the one-dimensional case, bigger
breathers eat up smaller ones, and finally only
one breather survives. However, note that,
depending on the initial conditions, the simulations do not always
lead to the coalescence into a single breather, because collisions
are more rare in two dimensions than in one. After the formation of a few
localized structures, one also observes the final relaxation to
equipartition, which is not shown in Fig. \ref{fran1}. This latter
is instead evident from the time evolution of $C_0(t)$, the
localization parameter, shown in Fig. \ref{cohlyap}(c): its
behavior is very similar to the one-dimensional case. Indeed, also
the largest finite time Lyapunov exponent behaves similarly (see
Fig. \ref{cohlyap}(d)).

\section{Spontaneous localization by edge cooling}
\label{Spontaneouslocalization}

Breathers play an important role in the non--equilibrium dynamics of the
FPU model. The relaxation to equipartition of the zone--boundary mode,
analyzed thus far in this paper, is one example. Another interesting
case in which breathers emerge spontaneously is when the lattice is
cooled at its boundaries~\cite{Aubry1,Aubry2,noi,reigada1,noichaos}. This
process may be thought of as modeling a non--equilibrium process where
energy exchange in the bulk is much slower than at the surface.
Although in this case the dynamics is dissipative, it turns out that
there is a deep connection with the original Hamiltonian model.

Specifically, when modelling this process, we add a dissipative term
$-\eta \dot u_n$ in the r.h.s. of Eq.~(\ref{sub}) when $n=1$ and $n=N$.
Similarly, in two dimensions, the same term is introduced at all edge
sites.  The parameter $\eta$ controls the strength of the coupling
with the external reservoirs at zero temperature. Since we are
interested in the exchange of energy between a finite system and the
environment, we shall consider either free--ends or fixed--ends
boundary conditions.

In a typical simulation, first an equilibrium micro-state is generated
by letting the Hamiltonian system ($\eta=0$) evolve for a sufficiently
long transient. Then, the dissipative dynamics ($\eta>0$) is started.
The initial condition for the Hamiltonian transient is assigned by
setting all relative displacements to zero and by drawing velocities
at random from a Gaussian distribution. The velocities are then
rescaled by a suitable factor to fix the desired value of the initial energy
$E(0)$ (see Ref.~\cite{noi} for the details).

\begin{figure}[ht]
\centering
\includegraphics[width=8.truecm,clip]{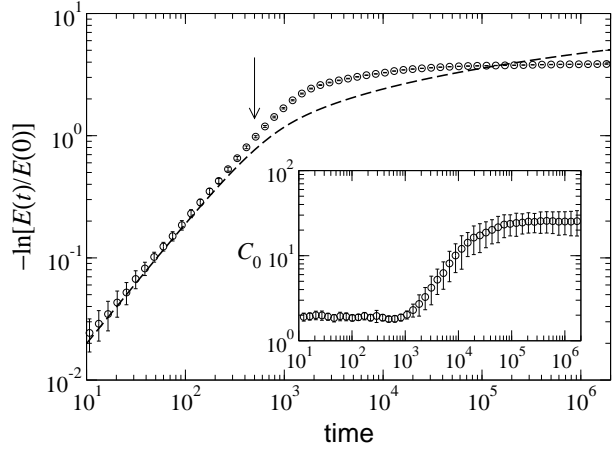}
\caption{\small
  Log--log plot of $-\log[E(t)/E(0)]$ versus time for an FPU chain
  with free--ends boundary conditions. In this representation, an
  exponential is a straight line with slope one. Symbols are the
  results of numerical simulations averaged over 20 initial
  conditions.  The dashed line is a plot of the theoretical
  result~(\ref{Totenfree})~\cite{noi} for a harmonic chain. The arrow
  indicates the crossover time $\tau_0$. Parameters are $N = 100$ and
  $\eta=0.1$.}
\label{f:Erel}
\end{figure}

The one-dimensional numerical simulations reveal that the dissipation
rate of the energy is dominated by two sequential effects, that
characterize the pathway to localization.  In the first stage of the
energy release process, the relaxation law undergoes a crossover from
the exponential $\exp[-t/\tau_{\scs 0}]$ to the power law $(t/\tau_{\scs
  0})^{-1/2}$, where $\tau_{\scs 0} = N/(2\eta)$ sets the shortest time
scale of the system (see Fig.~\ref{f:Erel})~\cite{noichaos}.
Asymptotically, energy reaches a plateau and, correspondingly, the
localization parameter $C_0(t)$ also saturates (see the inset of Fig.
\ref{f:Erel}).

The crossover at $\tau_{\scs 0}$ is a signature of the hierarchical
nature of the early stages of the process. In the harmonic
approximation, if one adds a small dissipation, the frequency $\omega$ of each
linear mode acquires an imaginary part $\gamma(\omega)$, which represents its
damping rate.  Initially, it is only the fastest mode which determines
the energy relaxation rate.  As time goes on, past $t\simeq\tau_{\scs 0}$,
it is the full spectrum of decay times of the linear modes that sets
the rules of energy relaxation.  Actually, it turns out that at this
stage, for small nonlinearities, the system behaves approximately as
its linear counterpart (see Fig.~\ref{f:Erel}). In particular, a
perturbative calculation to first order in $\gamma$ confirms
that~\cite{noichaos}
\begin{eqnarray}
\label{Totenfree}
\frac{E(t)}{E(0)}& \equiv & \int  e^{\ds -\gamma(\omega)t}\,
g(\omega)\, d \omega 
\simeq   \begin{cases}
e^{\ds -t/\tau_{\scs 0}} &
\text{for $t \ll \tau_{\scs 0}$} \\
\frac{\ds 1}{\ds \sqrt{2 \pi t/\tau_{\scs 0}}} &
\text{for $t \gg \tau_{\scs 0}$}
\end{cases} ,
\end{eqnarray}
where the density of states $g(\omega)$ is derived from the dispersion
relation $\omega(q)=2\sin(q/2)$, with $q=\pi k/N$ for free--ends boundary
conditions, and $q=\pi (k+1)/(N+1)$ for fixed--ends boundary conditions
($k=0,\dots N-1$).

The time range after the crossover coincides with the onset of
localization.  Now the dynamics is significantly affected by the
spontaneous appearance of breathers.  As the latter exhibit a very
weak interaction among themselves and with the boundaries, the energy
release undergoes a slowing down, thus freezing the system in a
quasi--stationary configuration far from thermal equilibrium.

\begin{figure}[ht]
\centering
\includegraphics[width=6. truecm,clip]{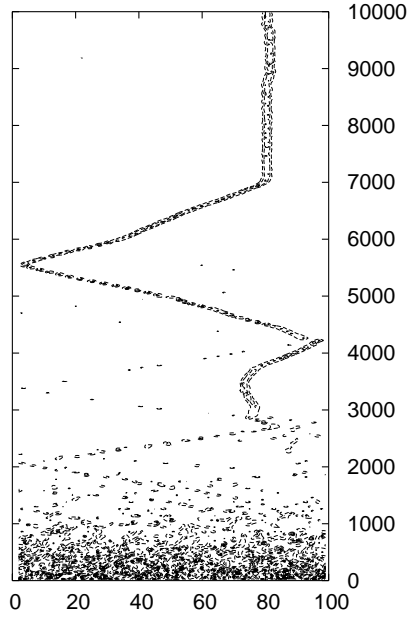}
\caption{\small  \label{f:cpl1D} Space--time contour plot of the
  site energies (\ref{localenergy}). Time flows up and the horizontal
  axis is the site index. Parameters are $N=100$, $\eta=0.1$ and
  $E(0)/N=1$. }
\end{figure}

Such residual state is characterized by the presence of one, highly
energetic localized object (possibly accompanied by a few much smaller
ones), which is mobile and alternates periods of rest and erratic
motion, as shown in Fig.~\ref{f:cpl1D}. This behaviour is reminiscent
of the chaotic breathers which emerge in the Hamiltonian system from
modulational instability of band--edge modes, discussed in
Section~\ref{localization1d} (see also Fig. \ref{greyscales}).

\begin{figure}[ht]
\includegraphics[width=8.5 truecm,clip]{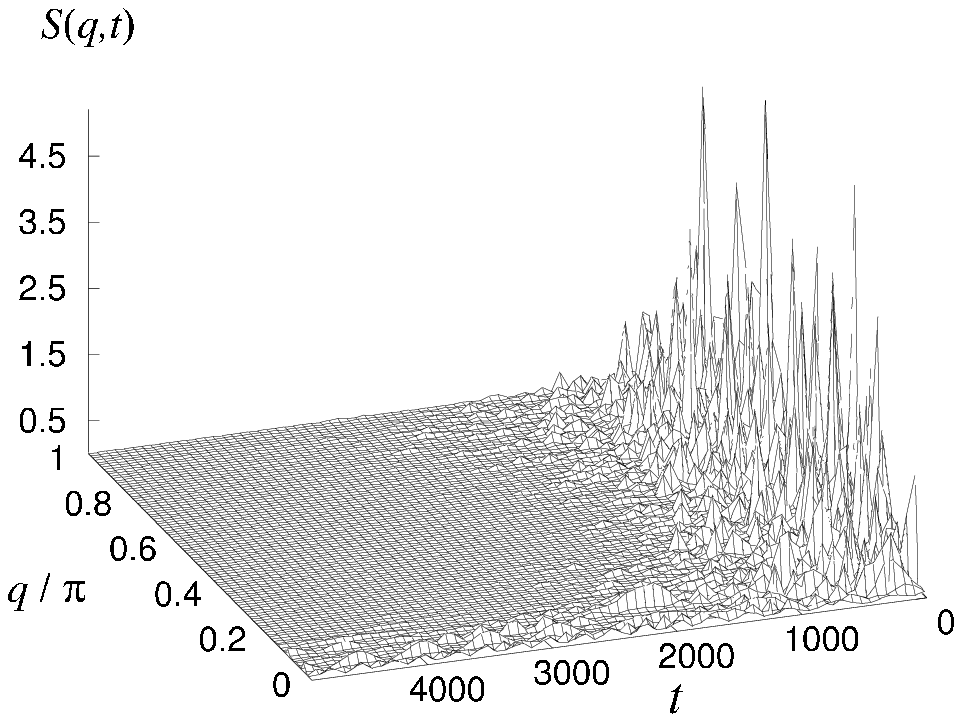}
\includegraphics[width=8.5 truecm,clip]{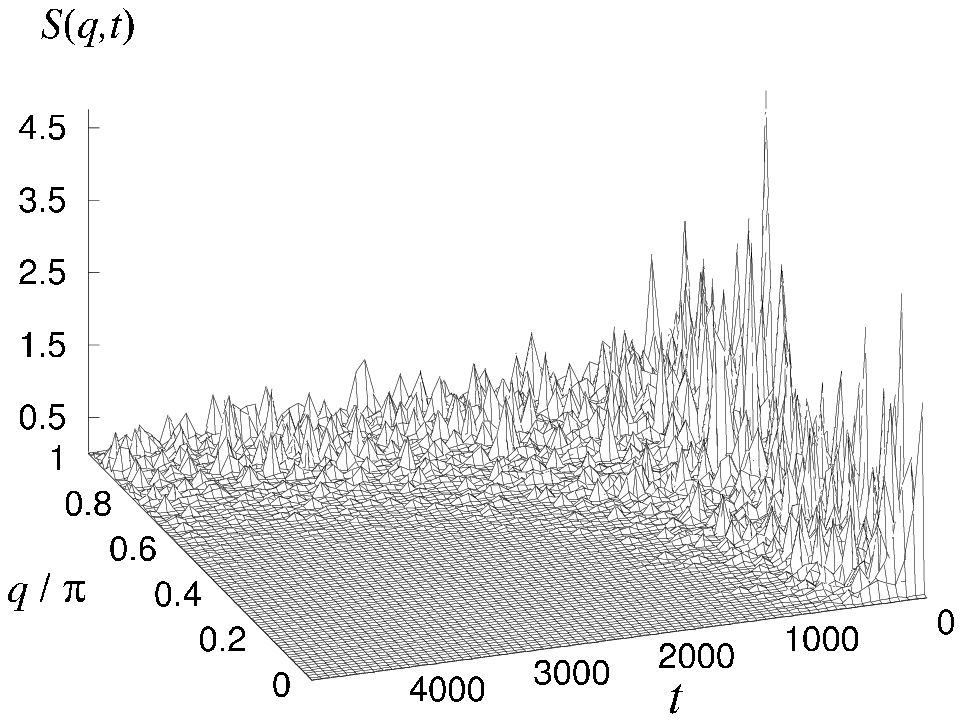}
\caption{\small Surface plot of the time--dependent spatial
  spectrum of particle velocities for the one-dimensional FPU lattice.
  (a) Fixed--ends boundary conditions. (b) Free--ends boundary
  conditions.  Parameters are $N=100$, $\eta=0.1$ and $E(0)/N=1$.}
\label{f:surfSQ}
\end{figure}

But what do we know of the mechanisms leading to spontaneous
localization in the presence of dissipation?   As
noticed above, there is evidence that it is the modulation instability
of short lattice waves that triggers the formation of localized
structures.  This hypothesis is supported by the observation that the
emergence of spatial patterns in the early stages of the relaxation is
intimately related to how dissipation acts on vibrational modes of
different wavelength.  In particular, if the system is swiftly enough
depleted of long--wavelength modes, the instability associated with
the band--edge waves may effectively trigger the process of
localization. A  key test for the above hypothesis is offered
by the nature of the boundary conditions.
In the case of free--ends, the modes of small wavenumber indeed
disappear very fast, in conjunction with the onset of spontaneous
localization.  On the contrary, in the presence of fixed boundaries
the former turn out to be as long--lived as the  modes with large
wavenumbers.
The corresponding numerical evidence is that hardly no localization is
observed in this case. Rather, the energy decays following exactly the
behavior of the harmonic chain~(\ref{Totenfree})~\cite{noi}.  This
scenario can be observed directly, by computing the time--dependent
spatial power spectrum $S(q,t)$ during the relaxation (see Fig.~\ref{f:surfSQ}).

\begin{figure}[ht]
\resizebox{0.5\textwidth}{!}{\includegraphics{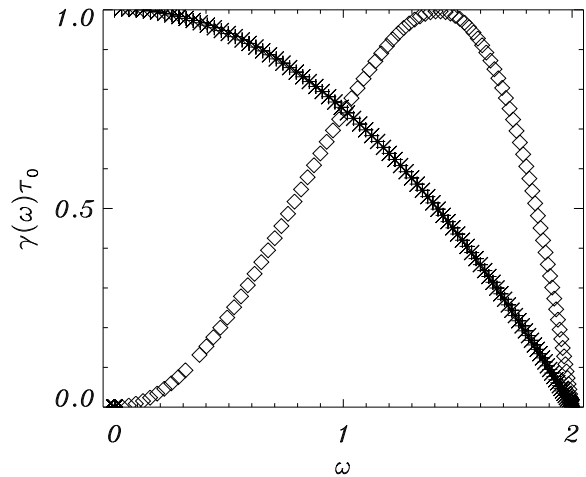}}

\caption{\small Damping rates $\gamma(\omega)$ for an  harmonic
  chain with $N=100$ and $\eta=0.1$.  Free--ends boundary conditions (stars) and
  fixed--ends boundary conditions (diamonds).}
\label{f:relspect}
\end{figure}

It is possible to get a quantitative confirmation of the above
hypothesis by calculating the exact relaxation spectrum $\gamma(\omega)$ of the
linearized system.  In the harmonic approximation, the equations of
motion may be written in matrix form as
\begin{equation}
\label{e:eqmotlin}
\dot{U} = \mathbb{A} \,  U,
\end{equation}
where  $U = (u_1,\dots,u_N,\dot{u}_1,\dots,\dot{u}_N)^T$ is a $2N$ column vector, and
$\mathbb{A}$ is the matrix
\begin{equation}
\label{Amatr}
\mathbb{A} = \left(
             \begin{BMAT}(e)[2pt,1cm,1cm]{c.c}{c.c} 0 & \mathbb{I}_N \\
             K & -\eta B \end{BMAT}
             \right) \quad .
\end{equation}
The tri--diagonal matrix of force constants $K_{nm}$ also contains the
information on the type of boundary conditions, whereas the matrix
$B_{nm}=\delta_{nm}(\delta_{m1}+\delta_{mN})$ describes the coupling with the
environment.  The spectrum of damping rates can be calculated by
straight diagonalization of matrix $\mathbb{A}$. In
Fig.~\ref{f:relspect} we plot  $\gamma(\omega)$, where
$\gamma$ is the opposite of the real part of an eigenvalue of $\mathbb{A}$
and $\omega$ its imaginary part.  This representation of the relaxation
spectrum is preferable with respect to the one in terms of wave-numbers, since the
vibration frequencies are shifted as a consequence of the damping.

This calculation confirms that the free--ends and fixed--ends systems
display considerably different behaviors. In the former case, the
least damped modes are the short--wavelength ones ($\omega\approx2$, the
band--edge frequency), the smallest damping constant being
$\gamma(2)\approx\pi^2\eta/2 N^3$, while the most damped modes are the ones in the
vicinity of $\omega\sim1/N$.  On the contrary, for fixed ends, the most damped
modes are those around the band center ($\omega\sim\sqrt{2}$) while both short-- and
long--wavelength waves dissipate very weakly ($\gamma\sim1/N^3)$.

The above analysis helps understanding why spontaneous localization is
strongly inhibited if the system is trapped between rigid walls, thus
in parallel unveiling the role of modulational instability in the
process.

\begin{figure}[ht]
\centering
\includegraphics[width=9.truecm, height=8. truecm,clip]{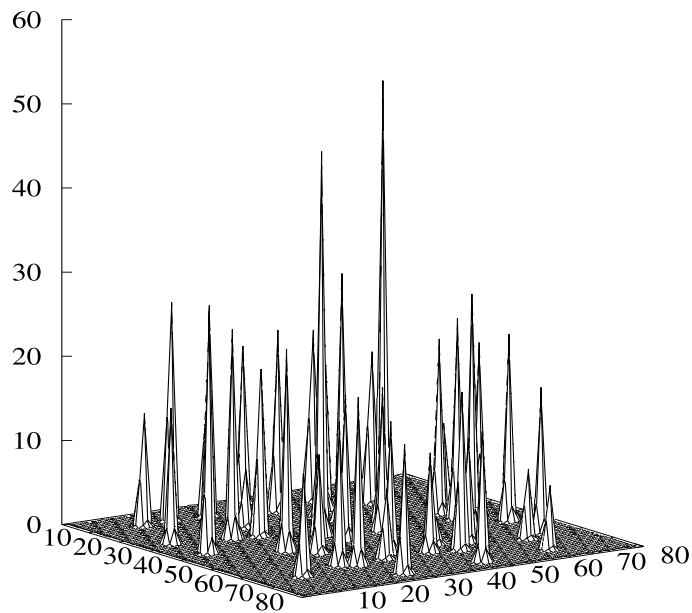}

\caption{\small
2D FPU lattice, site energies in the residual state state.
Parameters are: $N=80$, $\eta=0.1$,  $E(0)/N=1$.}
\label{f:2Dplot}
\end{figure}

We have also performed similar numerical simulations for the
two-dimensional FPU model~(\ref{2Dfpu}), with dissipation added at the
edges and free--ends boundary conditions.  Remarkably, the asymptotic
scenario changes. The quasi--stationary state is now a static
collection of tightly--packed localized objects, arranged in a sort of
random lattice (see Fig.~\ref{f:2Dplot}). Moreover, it turns out that
spontaneous localization in two dimensions is a thermally--activated
phenomenon, described by an Arrhenius law for the average breather
density, where the parameter that controls the strength of thermal
fluctuations is the initial energy density $E(0)/N$~\cite{noichaos}.
The origin of this behaviour is that, in the two-dimensional FPU
system, discrete breathers may be excited only above a certain energy
threshold~\cite{Kladko}, as discussed in Section~\ref{modinst2D}.
Despite the different nature of the asymptotic state, the onset of
localization follows the same path in one and two
dimensions~\cite{noichaos}, well described the hierarchy of relaxation
times underlying Eq.~(\ref{Totenfree}). In particular, a crossover is
 observed from the exponential $\exp[-2t/\tau_{\scs 0}]$ to the
power law $(t/\tau_{\scs 0})^{-1}$.

\section{Conclusions}
\label{conclusions}

In this paper, we have presented a detailed analysis of the
zone-boundary mode modulational instability for the FPU lattice in
both one and higher dimensions. Formulas for the critical
amplitude have been derived analytically and compare very well
with numerics for all system sizes. We have extended to two
dimensions the study of the process which leads to the formation
of chaotic breathers. The physical picture is similar to the
one-dimensional case. Besides that, we make the bridge between
breathers created by modulational instability of plane waves and
the ones formed when extracting energy from the boundaries:
similarities and differences are highlighted.

All results on modulational instability of zone-boundary modes can be
straightforwardly extended to other initial modes and, correspondingly,
instability rates can be derived. This has already partially been done
in Ref.~\cite{Paladin} and compares very well with the numerical
results by Yoshimura~\cite{yoshi2}. This author has recently
reanalyzed the problem~\cite{yoshi3} to determine the growth rates
for generic nonlinearities in the high energy region, obtaining exact
results based on Mathieu's equation.

Going to many--modes initial excitations, it has been remarked that
instability thresholds depend on relative amplitudes and not only on
the total energy~\cite{chechin2004}. Although this makes the study of
the problem extremely involved, we believe that a detailed study of
some selected group of modes, which play some special role in FPU
dynamics, could be interesting. The method discussed in this paper could
be adapted to treat this problem. Historically, the first study is in
the paper by Bivins, Metropolis and Pasta himself~\cite{bivins}, where
the authors tackle  the problem by studying numerically the
instabilities of coupled Mathieu's equations.

The study we have presented in this paper of the two-dimensional FPU
lattice is extremely preliminary and further analyses are needed. In
particular, the full process of relaxation to energy equipartion and
the associated time scales have not been studied at all. Preliminary
results on the relaxation process in two dimensions from low frequency
initial states seem to indicate a faster evolution to
equipartition~\cite{benettin}.  A similar analysis for high
frequencies remains to be performed.

In one-dimensional studies, a connection between the average
modulation instability rates and the Lyapunov exponents has been
suggested~\cite{DRT,Paladin}. Recently~\cite{franzozi}, high frequency
exact solutions have been used in the context of a differential
geometric approach~\cite{pettinilyap} to obtain accurate estimates of the
largest Lyapunov exponent. Similar studies could be performed for the
two-dimensional FPU lattice and the corresponding scaling laws with
respect to energy density could be obtained.

The study we have reported in the last Section about lattices that are
cooled at the boundaries points out the similarity of the localized
objects obtained in the long time limit with chaotic breathers.
However, this resemblance, although convincing, is only qualitative.
Quantitative studies on the comparison of these breathers with the
chaotic ones obtained from modulational instability should be performed.

\bigskip {\bf Acknowledgement:} First, we would like to thank D. K.
Campbell, P. Rosenau and G. Zaslavsky for the opportunity they gave us
to contribute to this Chaos issue celebrating the anniversary of the
FPU experiment.  Then, we express our gratitude to all our
collaborators in this field: J.  Barr{\'e}, M. Cl{\'e}ment, T. Cretegny, S.
Lepri, R.  Livi, P. Poggi, A. Torcini. We also thank N.  J. Zabusky
for useful exchanges of informations and Zhanyu Sun for an important
comment. This work is part of the contract COFIN03 of the Italian MIUR
{\it Order and chaos in nonlinear extended systems}.  R.Kh. is
supported by NATO reintegration grant No.  FEL.RIG.980767.  S.R.
thanks ENS Lyon for hospitality and financial support.

\end{document}